\documentclass[pra,aps,twocolumn]{revtex4}
\usepackage[12pt]{xy}
\usepackage{graphicx}
\usepackage{color}
\usepackage{amsmath}
\usepackage{slashbox}

\begin{document}

\title{$\chi^{(3)}$ non-Gaussian state generation for light using a
  trapped ion}

\author{Magdalena Stobi\'nska}
\email{mstobinska@optik.uni-erlangen.de} \affiliation{Institute f\"ur
  Optik, Information und Photonik, Max-Planck Forschungsgruppe,
  Universit\"at Erlangen-N\"urnberg,\\ G\"unter-Scharowsky-Str.\ 1,
  Bau 24, 91058, Germany}

\author{G. J. Milburn}
\affiliation{Centre for Quantum Computer Technology and School of
  Physical Sciences, The University of Queensland, St Lucia,
  Queensland 4072, Australia}

\author{Krzysztof W\'odkiewicz}
\email{wodkiew@fuw.edu.pl}
\affiliation{Instytut Fizyki Teoretycznej,
  Uniwersytet Warszawski, Warszawa 00--681, Poland}
\affiliation{Department of Physics and Astronomy, University of New
  Mexico, Albuquerque, NM~87131-1156, USA}

\date{\today}

\begin{abstract}

  According to the Gottesmann-Knill theorem the non-Gaussian states
  are necessary component for a nontrivial quantum computation.  We
  show two efficient and deterministic methods of $\chi^{(3)}$
  non-Gaussian state generation for a cavity mode using a single
  trapped ion. Both require ion motional state transfer to the cavity
  field. The methods are experimentally feasible. The first is based on
  the well-known protocol for an ion finite motional superposition
  state generation. It allows for an arbitrary good approximation of
  $\chi^{(3)}$ non-Gaussian states. We give criteria based on the
  Wigner function which quantify the error resulting from the
  approximation. The second and novel method enables an exact
  non-Gaussian state generation using one laser pulse only.
\end{abstract}

\maketitle

\section{Introduction}

The Gottesmann-Knill theorem for continuous variables (CV) states that
quantum computing based only on components described by one- or
two-mode quadratic Hamiltonians, Gaussian states input, and
measurements on canonical variables can be efficiently simulated by a
classical computation.  That means, although some of the algorithms
are of fundamentally quantum nature, they do not provide any speedup
over classical processes~\cite{Bartlett2002}.  It has also been proved
that construction of a CV universal quantum computer for
transformations that are polynomial in those variables requires cubic
or higher non-linear operations~\cite{Lloyd1999}. Therefore,
investigation of the non-Gaussian transformations and states
generation is crucial for a nontrivial quantum computation.  More
recently it has been noted that Kerr-like nonlinearities, in a variety
of systems, enable high precision quantum metrology that would
otherwise require entanglement to achieve it~\cite{caves}.

Over the last decade ion trap experiments have led the emerging
technologies of coherent quantum control, especially in quantum
information theory~\cite{Blatt2005} and quantum
computation~\cite{Ekert1996}.  Those applications require efficient
creation and precise manipulation of both electronic and motional
trapped ion state.  Various theoretical proposals on how to produce
nonclassical arbitrary states of ion motion have been discussed.  In
experiment, Fock number states~\cite{Roos1999}, coherent
states~\cite{Meekhof1996}, vacuum squeezed states~\cite{Heinzen1990},
and Schr\"odinger cat states~\cite{Monroe1996} have been realized.  In
this latter case, the state is an entangled state of the vibrational
and electronic degrees of freedom. On the contrary, in this paper we
give a deterministic way to prepare the vibrational degree of freedom
in a non-Gaussian state that is not entangled with the electronic
states. According to our knowledge, neither non-Gaussian state (other
than an entangled cat state) nor a superposition of more than two
coherent states has been observed so far.

In the case of photons there is no practical method of non-Gaussian state
generation so far. Efforts have been made to explore a class of
the $\chi^{(3)}$ non-Gaussian states produced using a photon coherent
state $|\alpha\rangle$ interacting with $\chi^{(3)}$ Kerr nonlinearity
in an optical fiber
\begin{equation}
|\Psi(\alpha, \tau)\rangle = e^{-\frac{|\alpha|^2}{2}}
\sum_{n=0}^{\infty}\, \frac{\alpha^n}{\sqrt{n!}}\,
e^{i\frac{\tau}{2}n(n-1)} |n\rangle.
\label{eq:kerr}
\end{equation}
This class of non-Gaussian states, parametrized by the unitless
evolution parameter $\tau$~\cite{Tanas}, is the most popular one. The
state (\ref{eq:kerr}) is known also as the Kerr state.  In general,
this is a highly nonclassical state and after a certain time of
evolution $\tau$ in the fiber its Wigner function would take negative
values in the phase space \cite{PRA}.  However, the nonlinearity in a
fiber, or any other experimentally achievable Kerr medium, is too
small, $\chi^{(3)} \simeq 10^{-22} \;
\frac{\mathrm{m^2}}{\mathrm{V^2}}$, to reach a highly nonlinear regime
and thus produce the negativity in an experimentally reasonable time,
before it is destroyed by dissipation \cite{GJM-CAH}.  Although
microstructured fibers seem to be more promising with $\chi^{(3)}
\simeq 10^{-16}\; \frac{\mathrm{m^2}}{\mathrm{V^2}}$, their length
does not exceed $1\,\mathrm{m}$ with current technology.

The most known examples of the Kerr state are the cat states $e^{-
  i\pi/4}|i \alpha\rangle + e^{i\pi/4} |- i\alpha\rangle$
corresponding to $\tau=\pi$, which have been found useful for studies
of quantum decoherence and quantum-classical boundary~\cite{Brune1996,
  Myatt2000}.  The larger cat states for which the two components $|i
\alpha\rangle$ and $|-i \alpha\rangle$ are almost orthogonal ($\alpha
> 1.5$) find their application in quantum information
processing~\cite{Ralph2003} and quantum
computation~\cite{Jeong,Jeong2}.

Recently, there has been introduced a probabilistic method of
non-Gaussian state generation relying on a conditional photon
subtraction from a squeezed vacuum
state~\cite{Grangier2006,Polzik2006,Furusawa2007,Ourjoumtsev}. Such a
state is a good approximation for the cat state if the amplitude is
small $\alpha <1$.  However, it is neither practical to subtract more
than two photons in the experiment nor to produce a state with
$\alpha>1.6$~\cite{Ourjoumtsev}.  Moreover, the state is not pure.

Another approach is suggested by recent achievements in the the Polzik
group~\cite{Polzik} based on their demonstration of teleportation of a
quantum state from optical to matter degrees of freedom.  If this
process was reversed, so that a non-Gaussian state is either
teleported or mapped from matter to light, optical Kerr states might
be robustly generated and thus overcome obstacles which are met with
in trying to generate optical non-Gaussian states directly. Of course
this requires us to demonstrate a way to make non-Gaussian states in
matter degrees of freedom.  The proposals for transferring an arbitrary
motional quantum state of an atom to a cavity field already exist
\cite{Zeng1994,Parkins1999,Pope2004}. The effort to integrate ion
traps and optical fields~\cite{ion-cavity} might offer a path based on
the extraordinary level of coherent quantum control one has over the
vibrational degrees of freedom for trapped ions.
 
In this paper we discuss two efficient and deterministic methods of
$\chi^{(3)}$ non-Gaussian state generation for light using a single
trapped ion. Both methods require ion motional state transfer to the
cavity mode. The first method is based on the well-known
protocol~\cite{Gardiner1997} for an ion finite motional superposition
state generation. It allows to produce the $\chi^{(3)}$ non-Gaussian
states with arbitrary good approximation. We give criteria, based on
Wigner function comparison and its measurement precision, which
quantify the error resulting from the approximation.  The second
method is novel and it enables an exact non-Gaussian state generation
using one laser pulse only.  We point out that a Wigner function
measurement of ion motional state can be performed using currently
available technology and already existing experimental schemas.  We
also suggest a quantum metrology application, based on the work of
Caves and co workers~\cite{caves}.

This paper is organized as follows. In section \ref{sec1} we show
that, applying a well-known protocol, one can produce an approximated
$\chi^{(3)}$ non-Gaussian state of motion for a trapped ion.  The
method relies on using a series of laser pulses to couple electronic
and vibrational degrees of freedom to effect the desired state
preparation for the motional degree of freedom.  We give the criteria
for quantifying the extent to which the prepared state approximates
the desired non-Gaussian state, and discuss the technical limitations
of the method.  In section \ref{sec2} we present an alternative, and
novel, method for generation of an exact $\chi^{(3)}$ non-Gaussian
state.  We also discuss the range of application of this method.  We
finish the paper with conclusions and a brief discussion of possible
applications to quantum metrology.

\section{Series of laser pulses method}
\label{sec1}

An ion in a Paul trap~\cite{Leibfried2003} may be prepared in an
arbitrary state of the form
\begin{equation}
|\Psi\rangle = \delta|\Psi^e\rangle |e\rangle + \beta|\Psi^g \rangle |g\rangle,
\end{equation}
using the method proposed in
\cite{Gardiner1997, Kneer1998}, where
\begin{equation}
|\Psi^e\rangle = \sum_{n=0}^{M} w_n^e |n\rangle,\quad |\Psi^g\rangle =
 \sum_{n=0}^{M} w_n^g |n\rangle  
\end{equation}
\noindent
are finite superpositions of ion motional states ($M<\infty$),
$|n\rangle$ is a Fock state of a harmonic oscillator potential in the
trap, $|g\rangle$ and $|e\rangle$ are the ion electronic ground and
excited states respectively.  Parameters $\delta$ and $\beta$ are
complex numbers obeying the normalization constraint
$|\delta|^2+|\beta|^2=1$ and are set at the start of the experiment. 

The method is based on applying a series of alternating laser pulses
tuned, first to the carrier, and then the red sideband, transition of
the trapped ion.  It works both within and beyond the Lamb-Dicke
regime.  The ion is initially prepared in its electronic and vibronic
(motional) ground state $|0,g\rangle$.  Adjusting the Rabi frequencies
and duration of each pulse properly, one could achieve $w_n^e = w_n^g
= w_n$ equal to
\begin{equation}
w_n =\frac{1}{\sqrt{\sum_{k=0}^M \frac{|\alpha|^{2k}}{k!}}}
\frac{\alpha^n}{\sqrt{n!}}\, e^{i\frac{\tau}{2}n(n-1)}.
\end{equation}
These coefficients correspond to the coefficients of a quantum
non-Gaussian state resulting from the unitary evolution of a self-Kerr
interaction (\ref{eq:kerr}) decomposed in the Fock basis, up to the
normalization factor (which results from the fact that we cut off the
infinite sum in (\ref{eq:kerr}) and take into account only the first
$M$ terms).

This method produces an approximated $\chi^{(3)}$ non-Gaussian state
$|\Psi^{(M)}(\alpha,\tau)\rangle$ corresponding to the state reached
via a Kerr interaction with an arbitrary value of evolution parameter
$\tau$.  Furthermore the effective value of $\tau$ can be much larger
than can be achieved via unitary interaction under a realistic optical
Kerr interaction were the bosonic degree of freedom an optical field
mode.  A proper choice of the cut-off value $M$ enables one to
approximate the state (\ref{eq:kerr}) arbitrarily closely. If we set
$\delta=0$ for simplicity, the ion state is given by
\begin{multline}
|\Psi^{(M)}(\alpha,\tau)\rangle = \frac{1}{\sqrt{\sum_{k=0}^M
    \frac{|\alpha|^{2k}}{k!}}} \sum_{n=0}^{M}\,
\frac{\alpha^n}{\sqrt{n!}}\, e^{i\frac{\tau}{2}n(n-1)} |n\rangle
|g\rangle.
\label{eq:approx_kerr}
\end{multline}
The number of required pulses for preparing the state
(\ref{eq:approx_kerr}) is equal to $2M$
\begin{equation}
|\Psi^{(M)}(\alpha,\tau)\rangle = R_M C_{M-1} \cdot ... \cdot R_1 C_0
|0,g\rangle,
\end{equation}
where $C_j$ and $R_j$ denote a carrier and a red sideband laser pulse
respectively. Therefore $M$ should be as small as possible
\cite{number_of_pulses}.

\begin{figure}
\scalebox{0.8}{\includegraphics{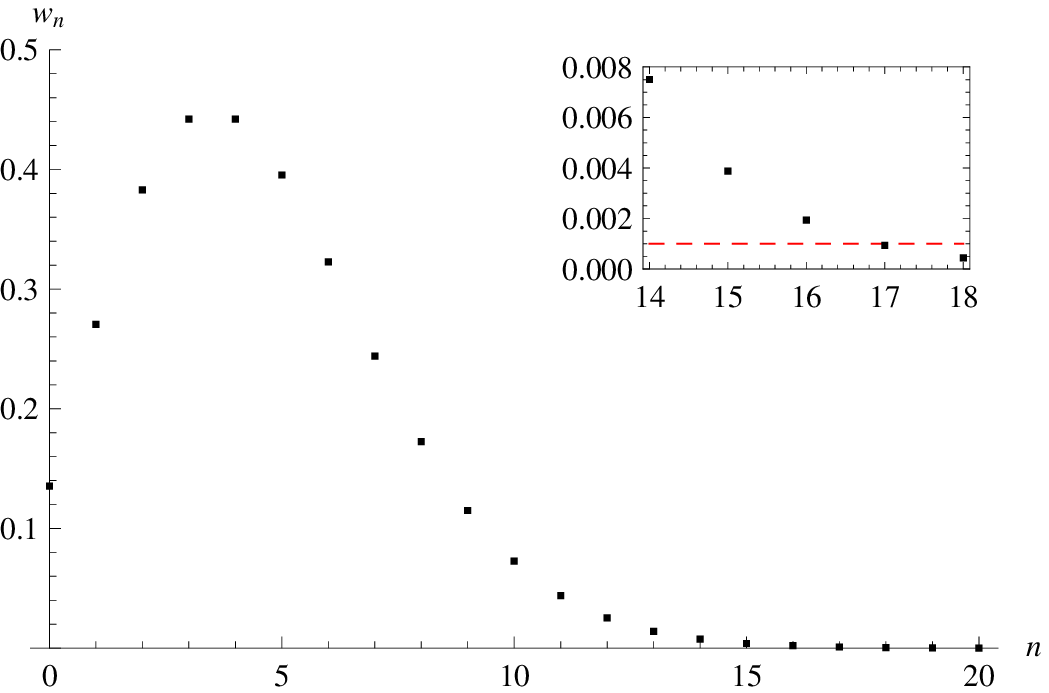}}
\scalebox{0.8}{\includegraphics{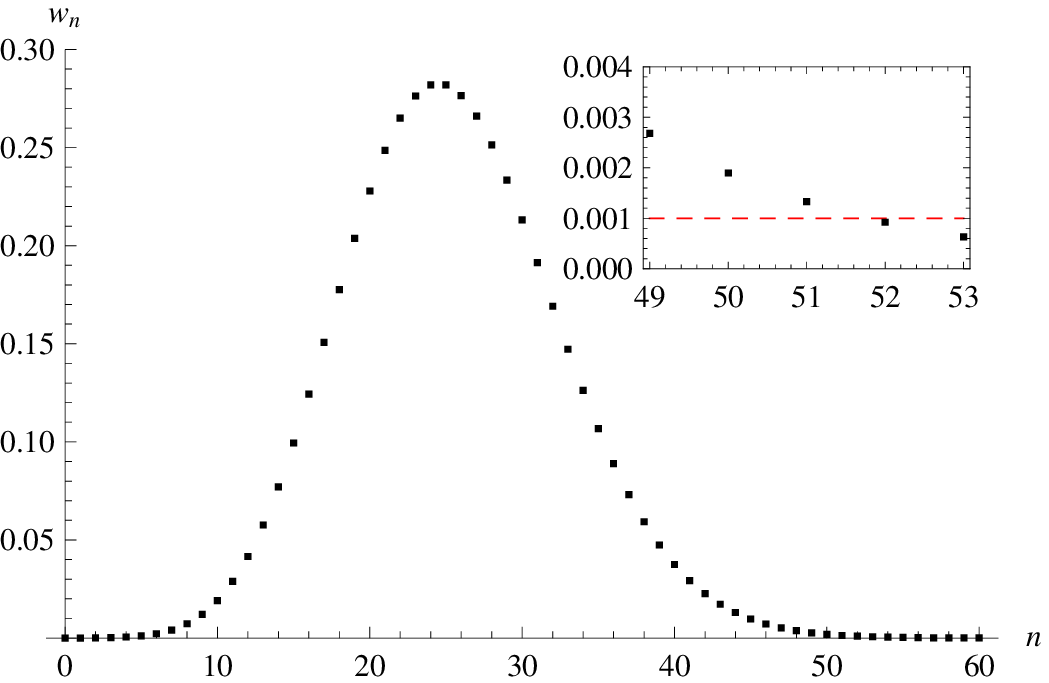}}
\caption{The coefficients $w_n$ of a $\chi^{(3)}$ non-Gaussian state
  decomposition in a Fock basis evaluated for $\alpha=2$ -- the top
  figure and $\alpha=5$ -- the bottom figure. The dots above the red
  dashed line take values greater or equal to $10^{-3}$. If $\alpha=2$
  ($\alpha=5$) the significant coefficients range from $w_0$ ($w_5$)
  to $w_{16}$ ($w_{51}$). }
\label{Fig:significant_coeff_a1}
\end{figure}

The hint for existence of a good non-Gaussian state approximation
comes from a simple observation that only a finite amount of
coefficients $w_n$ contribute to the sum (\ref{eq:kerr}) significantly
and the exact number of them is strongly $\alpha$ dependent.  The
coefficients, evaluated for exemplary values of the amplitude,
$\alpha=2$ and $\alpha=5$, are depicted on
Fig. \ref{Fig:significant_coeff_a1}.

For a given value of an amplitude $\alpha$ the choice of the cut-off
value $M$ is based on three criteria.  All criteria rely on comparing
 the Wigner function $W^{(M)}(\tau=2\pi,\gamma, \gamma^*)$ of an
approximated state (\ref{eq:approx_kerr}) with the untruncated one
$W(\tau=2\pi,\gamma, \gamma^*)$ evaluated for (\ref{eq:kerr}). Please
note that the comparison is made for $\tau=2\pi$.  We chose this
particular value of the evolution parameter because for this value the
original Wigner function is given by a simple analytic formula, which
can be computed directly.

The first criterion derives from an investigation of the approximated
Wigner function isolines. The smaller the value of $M$ the more the
function deviates from an ideal Gaussian function for $\tau=2\pi$: the
isolines are no longer circles.  The isolines of interest can be
chosen arbitrarily. We however chose $0.1$, $0.3$, $0.5$ and
calculate, for each isoline, the ratio between the most and the least
distant point with respect to the point $(\alpha,0)$ separately.  For
an ideal circle the ratio is always equal to $1$.

In order to take into account the interference resulting from the
approximate state, which takes the values around zero, we calculate
the maximal and average ratio of the difference
$\frac{\pi}{2}\left|W(\tau=2\pi,\gamma, \gamma^*) \right. -
\left. W^{(M)}(\tau=2\pi,\gamma, \gamma^*)\right|$ to the value of the
ideal Wigner function, for all points $\gamma$ in a phase space.
Therefore, the second criterion gives the maximal and average relative
error of the approximation respectively.

The third criterion reveals the percentage of points $\gamma$ in a
phase space for which $W^{(M)}(\tau=2\pi,\gamma, \gamma^*) =
W(\tau=2\pi,\gamma, \gamma^*)$ for a given precision.  We assume a
precision of order of either $\pm 10^{-2}$ or $\pm 10^{-3}$ to be good
enough,  since it relates to the accuracy of both the
Wigner function computer visualisation using the density plots and the
Wigner function reconstruction using quantum
tomography~\cite{Grangier_cat}.

The above criteria turn out to be more subtle than the well-known
fidelity $\mathcal{F}$ which measures the overlap of the approximated
and the original state in the phase space
\begin{equation}
\mathcal{F} = \left|\langle \Psi^{(M)}(\alpha,\tau)|
\Psi(\alpha,\tau)\rangle\right|^2 = e^{-2|\alpha|^2}
\left|\sum_{m=0}^{M}\frac{|\alpha|^{2m}}{m!}\right|^2.
\end{equation}
Obviously, the fidelity approaches unity in the limit of $M \to
\infty$.

We would like to point out that having the ion already prepared in
state (\ref{eq:approx_kerr}) one could measure the Wigner function of
its vibronic state using either the standard method of quantum
tomography \cite{Wallentowitz1995, Poyatos1996} or the method of
direct measurement developed in~\cite{Lutterbach1997}.  This would
allow for detailed investigation of $\chi^{(3)}$ non-Gaussian states,
which has never been verified experimentally so far.

The direct Wigner function measurement method is especially interesting.  
It relies on the fact that the Wigner function of a
displaced vibronic state of an ion is related to the probability of
finding the ion in the ground and the excited state. The probabilities
are measured by detecting a fluorescence signal.  Such a measurement
has been done for a light in a cavity with accuracy of $\pm
0.2$~\cite{direct_measurement_wigner}.

\subsection*{Example: $|\Psi^{(M)}(\alpha = 2, \tau)\rangle$ generation}

Estimation of the minimal number of laser pulses for
$|\Psi^{(M)}(\alpha = 2, \tau)\rangle$ generation, which approximates
the original state $|\Psi(\alpha = 2, \tau)\rangle$ for any value of
the evolution parameter $\tau$ well, requires comparison of the
approximated Wigner functions $W^{(M)}(\tau=2\pi,\gamma, \gamma^*)$
evaluated for a few different cut-off values $M$, with the original
Wigner function $W(\tau=2\pi,\gamma, \gamma^*)$.  The selection of the
best possible cut-off values is based on analysis of significant
coefficients $w_n$. For $\alpha=2$ only the coefficients ranging from
$w_0$ to $w_{16}$ are greater than $10^{-3}$. The coefficients
$w_{10}$ to $w_{13}$ are of order of $10^{-2}$. Therefore, we test $10
\le M \le 16$.

We next investigate the isolines. Table \ref{tabela_a2} shows the
ratios of the most to the least distant point, with respect to the
point of $(2, 0)$ for the isolines ($0.1$, $0.3$, $0.5$) of the
approximated Wigner functions $W^{(M)}(\tau=2\pi,\gamma, \gamma^*)$.
The minimal value of $M$ for which the ratios evaluated for
$W^{(M)}(\tau=2\pi,\gamma, \gamma^*)$ are equal to the ratios
evaluated for the original Gauss function using the grid (discretized
phase space) with a step equal to $\Delta \gamma = 0.04$ is equal to
$14$. It means that for the given grid and greater values of $M$, the
numerical simulations of the Wigner function will not differ.
Therefore, the value of $M=14$ could be regarded as an appropriate
cut-off value.

\begin{table*}[p]
\begin{center}
\begin{tabular}{c|c|c|c|c|c|c|c|c}

\backslashbox{Isoline}{M}&
          9             & 10            & 11            & 12        & 13        & 14        & 15 & 16 \\
\hline

0.1     & 1.33 (33\,\%) & 1.20 (20\,\%) & 1.10 (10\,\%) & 1.04 (--) & 1.01 (--) & 1.01 (--) & 1.00 (--) & 1.00 (--) \\

0.3     & 1.23 (23\,\%) & 1.10 (10\,\%) & 1.04 (--)     & 1.01 (--) & 1.00 (--) & 1.00 (--) & 1.01 (--) & 1.01 (--) \\

0.5     & 1.19 (19\,\%) & 1.08 (8\,\%)  & 1.01 (--)     & 1.04 (--) & 1.04 (--) & 1.04 (--) & 1.04 (--) & 1.04 (--)   

\end{tabular}
\end{center}
\caption{The ratios of the most to the least distant point, with
  respect to the point $(2, 0)$, for given isolines $0.1$, $0.3$,
  $0.5$, of the approximated Wigner function
  $W^{(M)}(\tau=2\pi,\gamma, \gamma^*)$ for $9 \le M\le 16$.}
\label{tabela_a2}
\end{table*}

\begin{table*}[p]
\begin{center}
\begin{tabular}{c|c|c|c|c|c|c|c|c}

\backslashbox{Accuracy}{M}&
            9      & 10     & 11     & 12     & 13     & 14    & 15 & 16 \\ \hline

$10^{-2}$ & 64\,\% & 76\,\% & 90\,\% & 99\,\% & 100\,\% & 100\,\% & 100\,\% & 100\,\%   \\

$10^{-3}$ & 26\,\% & 36\,\% & 44\,\% & 53\,\% & 65\,\%  & 80\,\%  & 97,\%   & 100\,\%

\end{tabular}
\end{center}
\caption{The number of points $\gamma$ in the phase space for which
  $W^{(M)}(\tau=2\pi,\gamma, \gamma^*) = W(\tau=2\pi,\gamma,
  \gamma^*)$ at a given accuracy for $9 \le M\le 16$.}
\label{tabela_a21}
\end{table*}

The series of $28$ laser pulses is experimentally feasible
\cite{number_of_pulses}.  However, decreasing the quality of
approximation only a little, one can diminish the number of pulses to
$20$. The ratios evaluated for $M=10$ ($M=9$) and isolines $0.1$,
$0.3$ and $0.5$ differ from unity of $20 \,\%$ ($33\,\%$), $10 \,\%$
($23\,\%$) and $8\,\%$ ($19\,\%$) respectively.  This is the minimal
cut-off value for which the average error equal to is below $1\,\%$
(see table \ref{tabela_a2_bledy}).  For $M=9$ the error is equal to
$1.67 \,\%$.  The maximal error is equal to $6.39 \,\%$ ($11.30
\,\%$).  Within a given precision $10^{-2}$ for $M=10$ ($M=9$) there
are around $76 \,\%$ ($64 \,\%$) points for which the numerically
obtained Wigner function values are equal to the values of the ideal
function. For the precision of $10^{-3}$ and $M=10$ ($M=9$) there are
$36 \,\%$ ($26 \,\%$) of such points.

\begin{table*}[p]
\begin{center}
\begin{tabular}{c|c|c|c|c|c|c|c|c}

\backslashbox{Error}{M}&
          9         & 10       & 11       & 12       & 13       & 14       & 15 & 16 \\ \hline

Average & 1.67\,\%  & 0.98\,\% & 0.55\,\% & 0.30\,\% & 0.15\,\% & 0.08\,\% & 0.04\,\% & 0.02\,\% \\

Maximal & 11.30\,\% & 6.39\,\% & 3.48\,\% & 1.84\,\% & 0.94\,\% & 0.47\,\% & 0.22\,\% & 0.11\,\%

\end{tabular}
\end{center}
\caption{The average and the maximal error estimated for $\alpha=2$ and
  for $9 \le M\le 16$.}
\label{tabela_a2_bledy}
\end{table*}

\begin{figure}
\begin{center}
  \scalebox{0.5}{\includegraphics{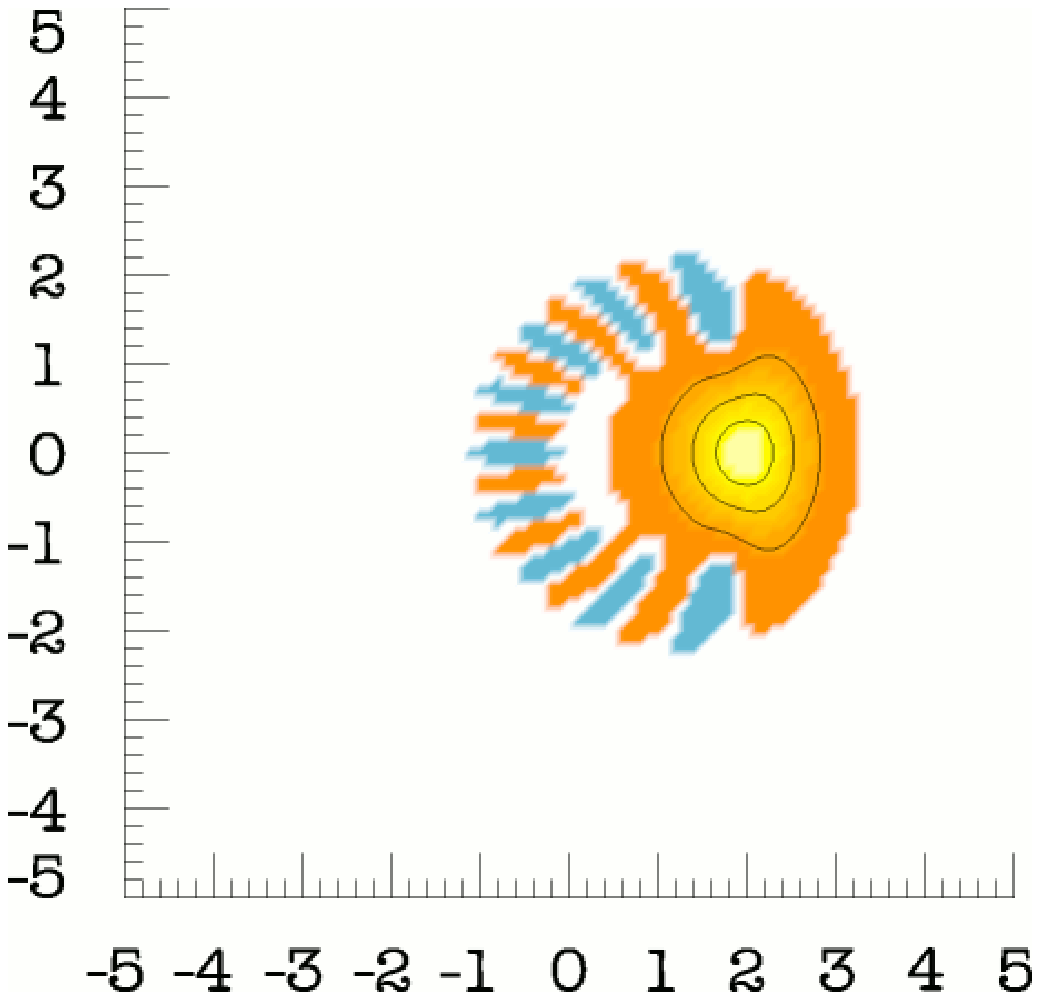}}\\[5mm]
  \scalebox{0.5}{\includegraphics{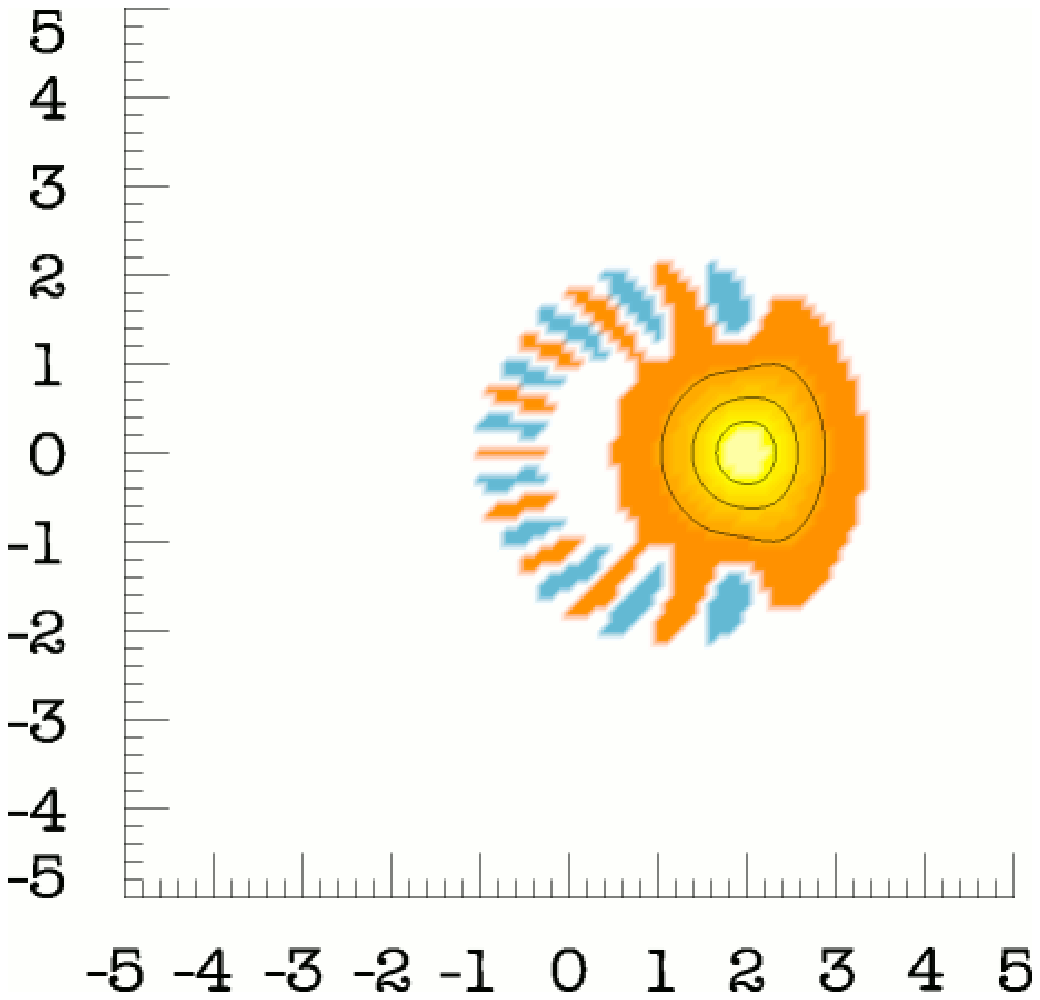}}\\[5mm]
  \scalebox{0.5}{\includegraphics{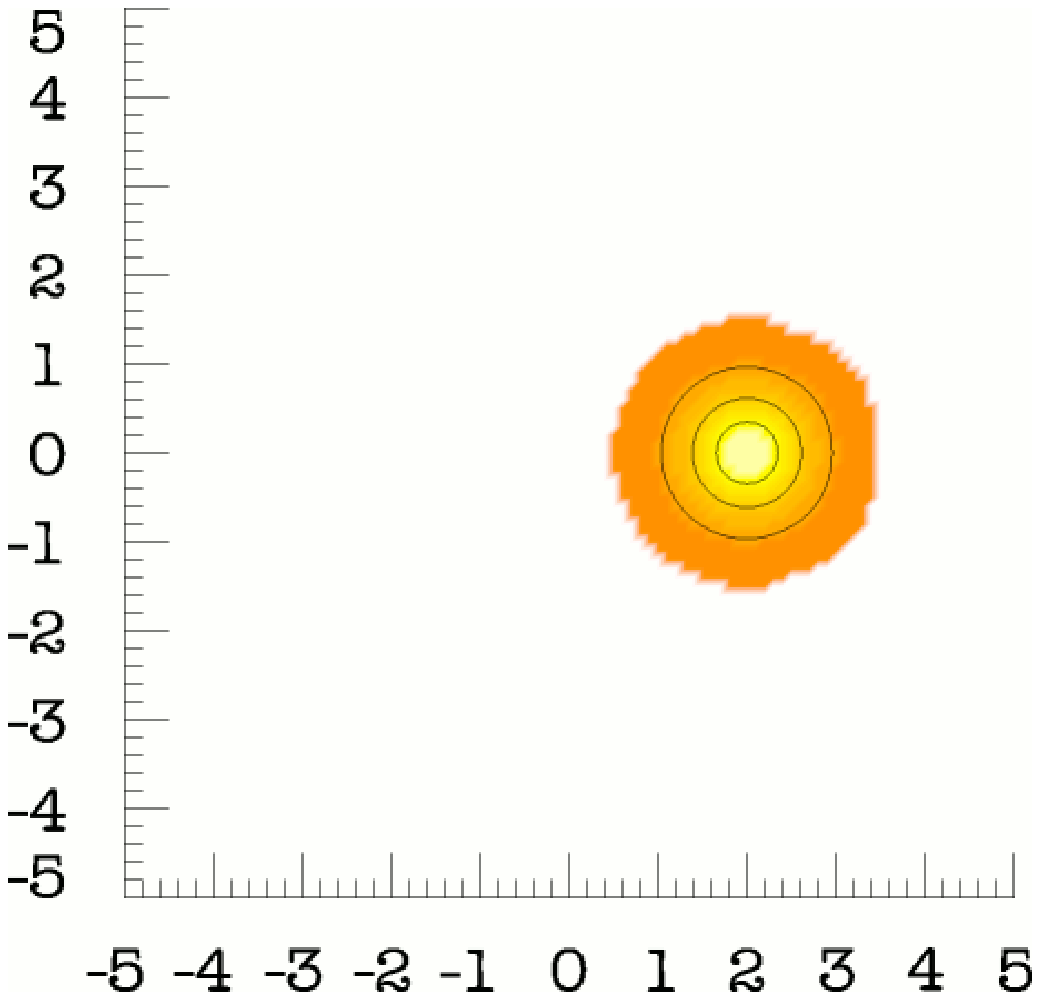}}\\[5mm]
  \scalebox{0.5}{\includegraphics{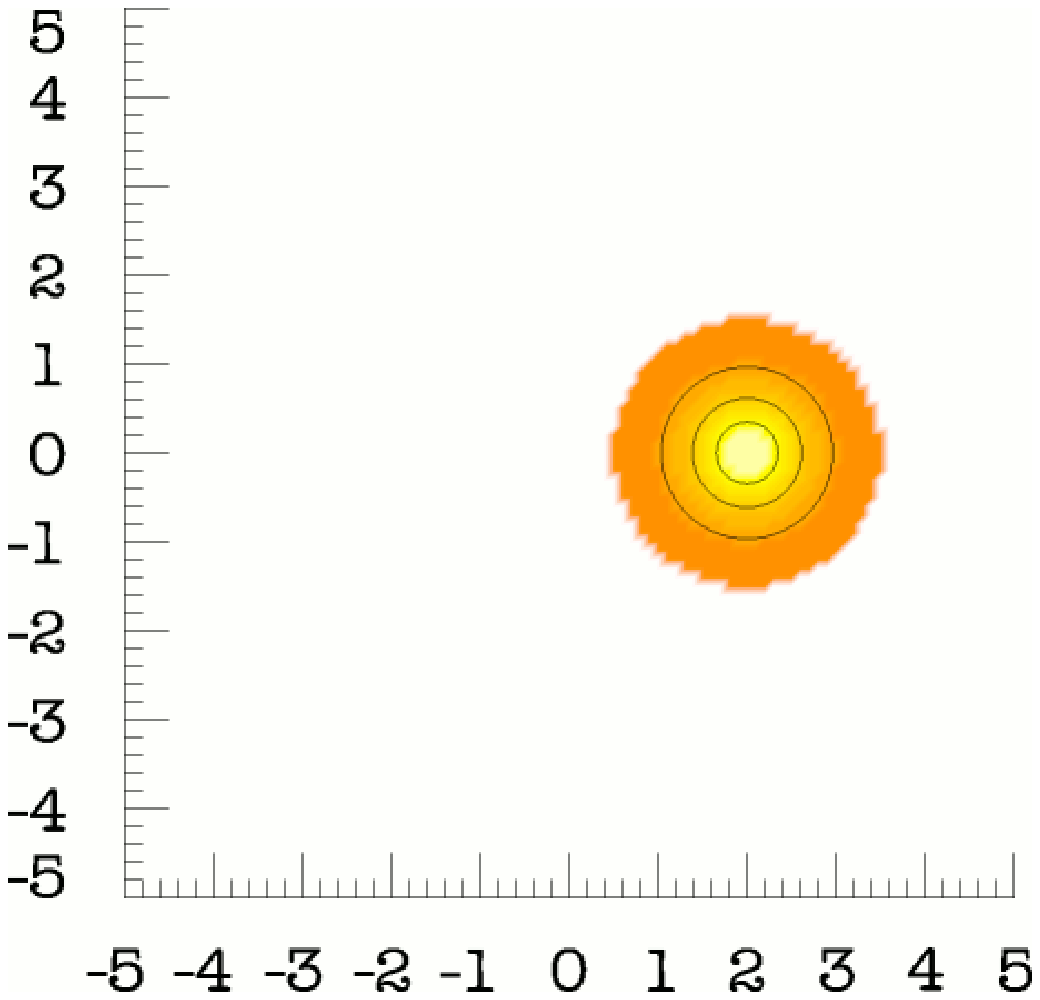}}\\[5mm]
  \scalebox{0.5}{\includegraphics{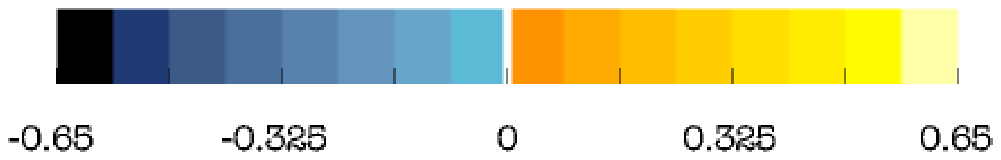}}
\end{center}
\caption{The Wigner function for an approximated $\chi^{(3)}$
  non-Gaussian state: $|\Psi^{(9)}(\alpha = 2,\tau=2\pi)\rangle$ --
  the top figure, $|\Psi^{(10)}(\alpha = 2,\tau=2\pi)\rangle$ -- the
  top middle figure, $|\Psi^{(14)}(\alpha = 2,\tau=2\pi)\rangle$ --
  the bottom middle figure, $|\Psi(\alpha = 2,\tau=2\pi)\rangle$ --
  the bottom figure.}
\label{Fig:wigner_a2_t2pi_compareM}
\end{figure}

These results show that $M=10$ is still a good approximation for the
Wigner function analysis and measurement.  The approximated functions
for $M=10$ and $M=14$ with marked isolines are depicted on
Fig. \ref{Fig:wigner_a2_t2pi_compareM}.  We also include the plot for
$M=9$ for comparison.

The fidelities evaluated for $|\Psi^{(9)}(\alpha = 2, \tau)\rangle$,
$|\Psi^{(10)}(\alpha = 2, \tau)\rangle$ and $|\Psi^{(14)}(\alpha = 2,
\tau)\rangle$ are equal to $\mathcal{F} = 0.9838$, $\mathcal{F} =
0.9943$, and $\mathcal{F} = 0.9999$ respectively.

Below we present the time of pulses duration and their phase required
to generate an exemplary compass state $|\Psi^{(10)}(\alpha = 2,\tau=
\pi/2)\rangle$, which is a superposition of four coherent states

\begin{center}
\begin{tabular}{r|r|r}
& \multicolumn{1}{c|}{$\varphi_{n}^R$}
& \multicolumn{1}{c}{$t_{n}^R$} \\ \hline
$R_{10}$& $\pi$  & $0.99\,\mathrm{ms}$ \\
$R_{9}$& $0.47$  & $0.39\,\mathrm{ms}$ \\
$R_{8}$& $7.23$  & $0.35\,\mathrm{ms}$ \\
$R_{7}$& $4.26$  & $0.44\,\mathrm{ms}$ \\
$R_{6}$& $5.00$  & $0.47\,\mathrm{ms}$ \\
$R_{5}$& $1.82$  & $0.55\,\mathrm{ms}$ \\
$R_{4}$& $2.21$  & $0.55\,\mathrm{ms}$ \\
$R_{3}$& $-1.30$ & $0.75\,\mathrm{ms}$ \\
$R_{2}$& $-0.95$ & $0.81\,\mathrm{ms}$ \\
$R_{1}$& $3.23$  & $1.37\,\mathrm{ms}$ 
\end{tabular}
\qquad
\begin{tabular}{r|r|r}
& \multicolumn{1}{c|}{$\varphi_{n}^C$}
& \multicolumn{1}{c}{$t_{n}^C$} \\ \hline
$C_{9}$& $0.00$  & $2.89\,\mu\mathrm{s}$ \\
$C_{8}$& $-0.83$ & $1.16\,\mu\mathrm{s}$ \\
$C_{7}$& $1.33$  & $1.30\,\mu\mathrm{s}$ \\
$C_{6}$& $2.41$  & $2.21\,\mu\mathrm{s}$ \\
$C_{5}$& $-0.05$ & $1.60\,\mu\mathrm{s}$ \\
$C_{4}$& $-0.86$ & $2.44\,\mu\mathrm{s}$ \\
$C_{3}$& $-2.97$ & $1.92\,\mu\mathrm{s}$ \\
$C_{2}$& $-3.93$ & $2.84\,\mu\mathrm{s}$ \\
$C_{1}$& $-0.09$ & $2.84\,\mu\mathrm{s}$ \\
$C_{0}$& $-4.19$ & $1.04\,\mu\mathrm{s}$
\end{tabular}
\end{center}

These quantities were evaluated for the following Rabi frequencies
$\Omega_C = 1 \, \mathrm{MHz}$, $\Omega_R = 100 \, \mathrm{kHz}$
corresponding to carrier transition and red sideband respectively, and
Lamb-Dicke parameter $\eta=0.02$.  The total time of state build-up is
equal to $t_c \simeq 6.7 \, \mathrm{ms}$.

\begin{figure}
\begin{center}
  \scalebox{0.5}{\includegraphics{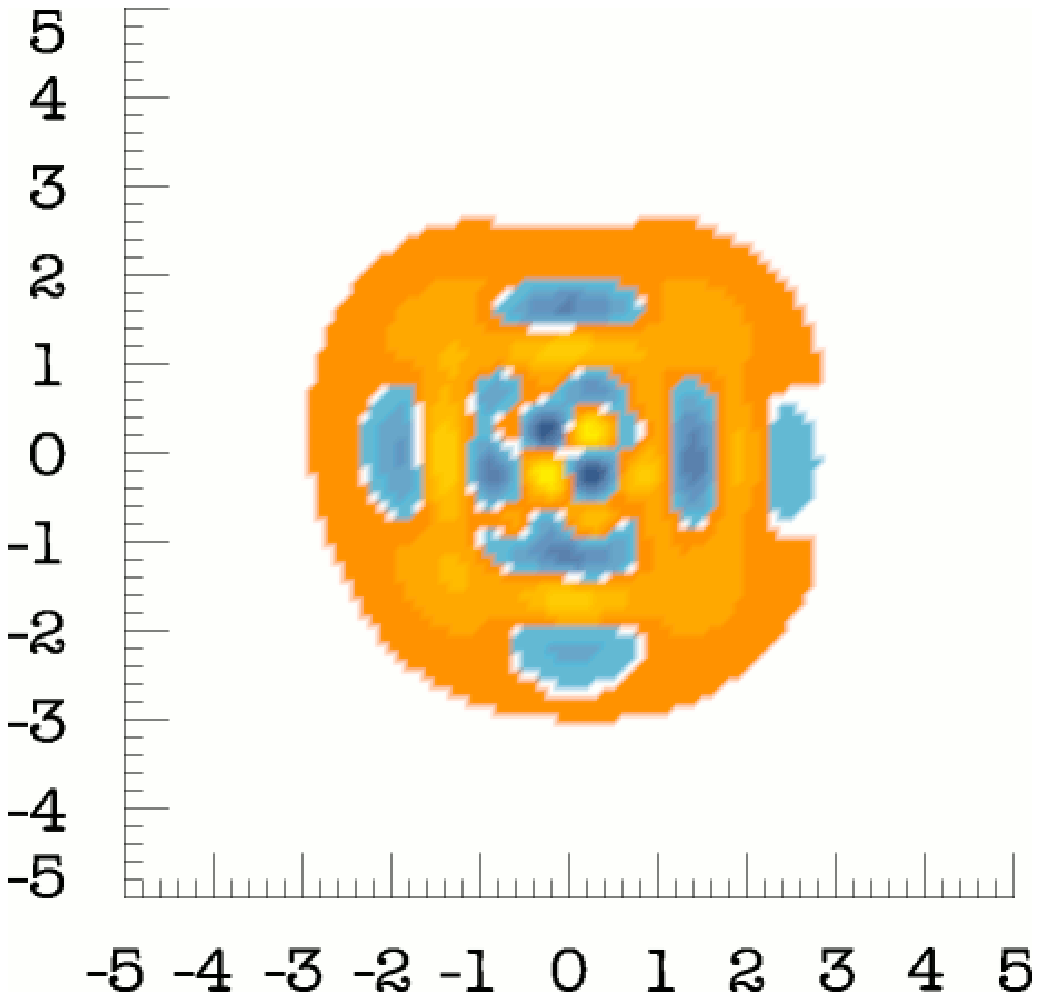}}\\[5mm]
  \scalebox{0.5}{\includegraphics{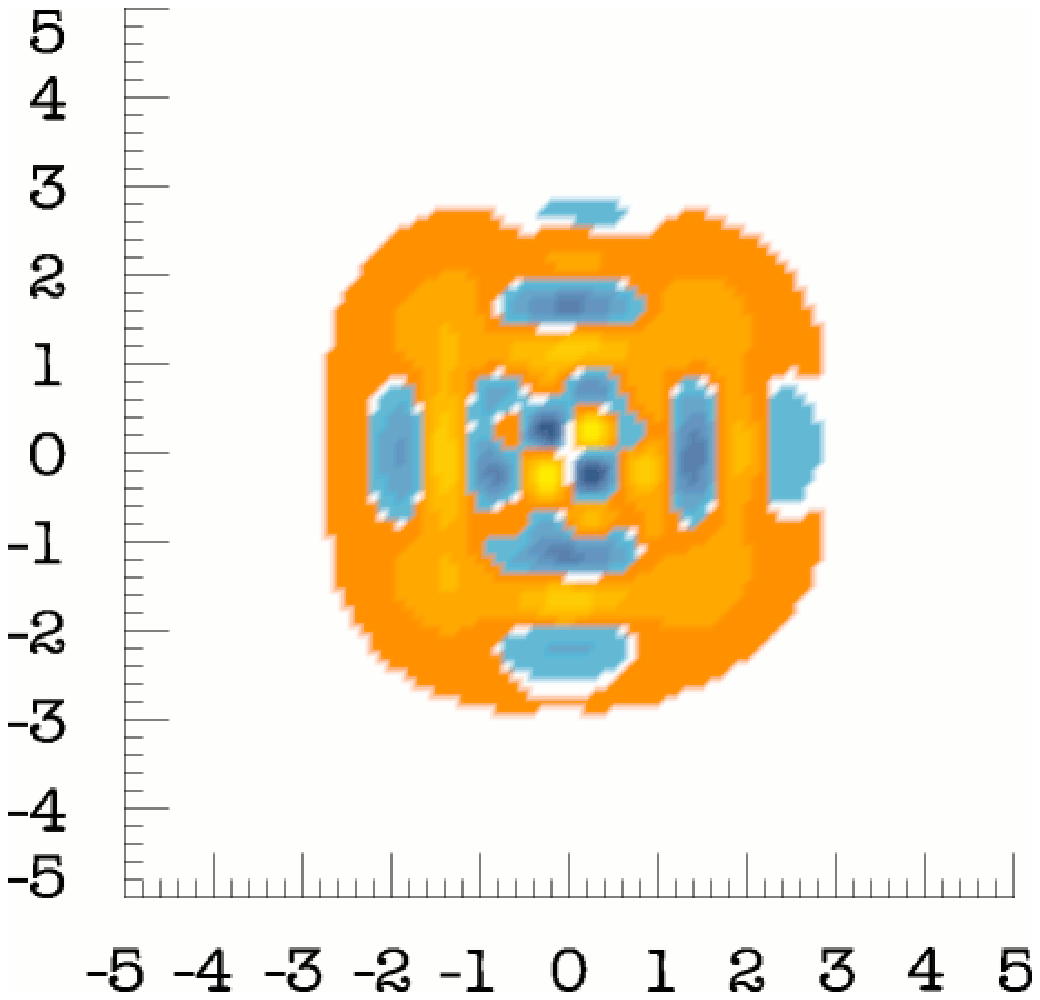}}\\[5mm]
  \scalebox{0.5}{\includegraphics{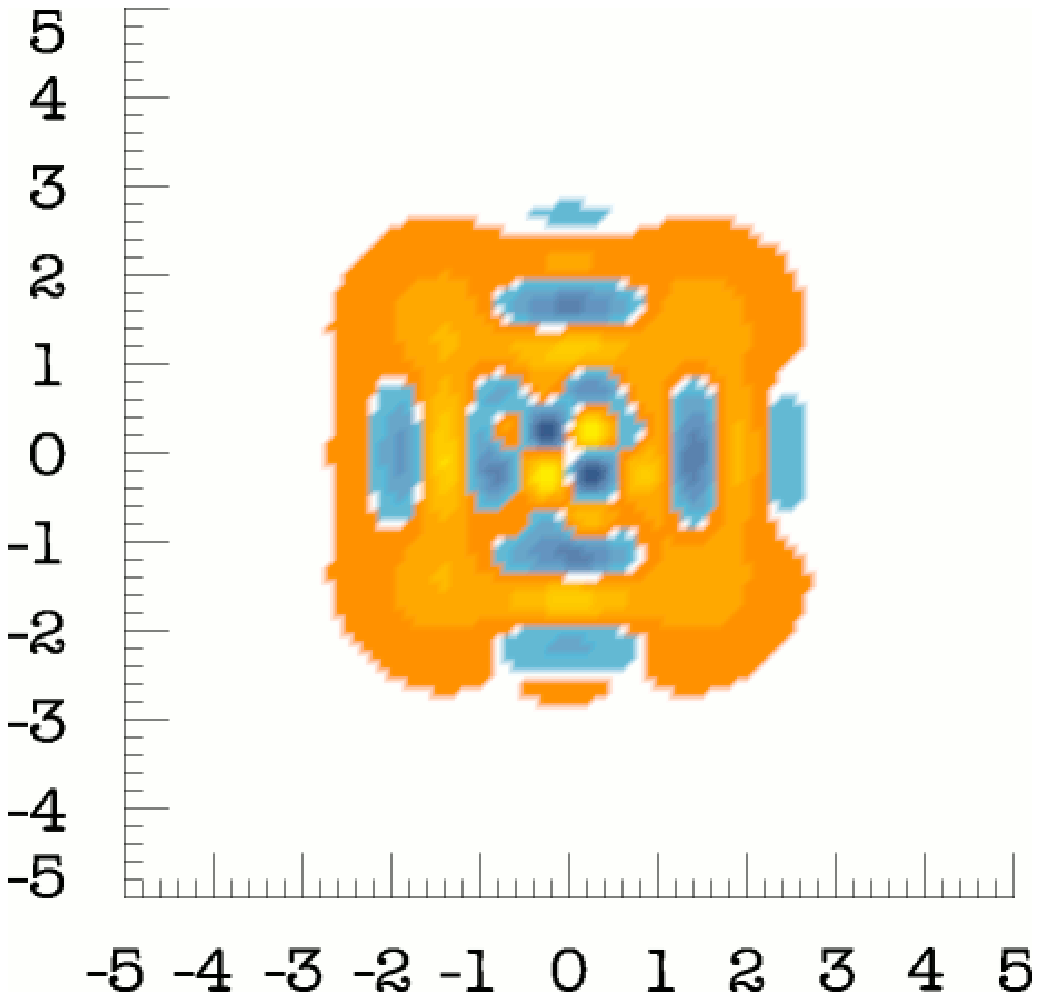}}\\[5mm]
  \scalebox{0.5}{\includegraphics{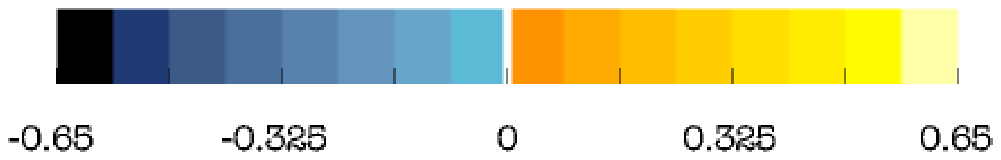}}
\end{center}
\caption{The Wigner function for an approximated $\chi^{(3)}$
  non-Gaussian state: $|\Psi^{(9)}(\alpha =
  2,\tau=\frac{\pi}{2})\rangle$ -- the top figure,
  $|\Psi^{(10)}(\alpha = 2,\tau=\frac{\pi}{2})\rangle$ -- the middle
  figure, $|\Psi(\alpha = 2,\tau=2\pi)\rangle$ -- the bottom figure.
}
\label{Fig:wigner_a2_t0.5pi_compareM}
\end{figure}

The Wigner function of the approximated compass state for $M=10$ is is
presented on Fig. \ref{Fig:wigner_a2_t0.5pi_compareM}.  We also
present the Wigner functions for $M=9$ and the original one for
$|\Psi(\alpha=2, \tau=\frac{\pi}{2})\rangle$ for comparison.

\subsection*{Technical limitations}

At the end of this section we estimate the maximum value of the
amplitude $\alpha$ for which the method works. The maximum value seems
to be $\alpha \simeq 2.3$.

This limitation does not result from the decoherence of the ion
state. Both the time of single pulse duration and the total duration
of state built-up remain within the coherence of the ion for the
vibronic state $190 \, \mathrm{ms}$. The coherence time for the
electronic state is equal to $1.4 \, \mathrm{ms}$.

The number of laser pulses is limited due to the finite binding energy
in the trap. The cut-off value $M$ corresponds to the maximal ion
excitation $|M\rangle$. The $|M=17\rangle$ is the upper limit for
trapping the ion so far~\cite{Blatt_disc}.

\section{One laser pulse method} \label{sec2}

The one laser pulse method of $\chi^{(3)}$ non-Gaussian state
generation requires one carrier resonance pulse applied to the ion
cooled down to a Lamb-Dicke regime. The vibronic state of the ion is
initially prepared in a coherent state $|\alpha\rangle$. The advantage
of this method over the first one is that it allows for generation of
the original Kerr state (\ref{eq:kerr}) without any
approximations.  However, not all values of the evolution parameter are
accessible: the pulse duration is limited due to the ion decoherence.

The interaction between an ion and a laser pulse of Rabi frequency
$\Omega$ is governed by the following Hamiltonian
\begin{equation}
H = \hbar \frac{\Omega}{2} \left\{\sigma^+
e^{i\eta(ae^{-i\nu}+a^{\dagger }e^{i\nu})} + h.c.\right\},
\end{equation}
where $\sigma^+$ is an electronic state rising operator, $a$ is a
vibronic state annihilation operator, $\nu$ is a trapping frequency,
and $\eta$ is a Lamb-Dicke parameter.  Using the expansion of the
exponens function into a Taylor series
\begin{equation}
e^{i\eta(ae^{-i\nu}+a^{\dagger }e^{i\nu})} = \sum_{k=0}^{L}
\frac{(i\eta)^k}{k!}(ae^{-i\nu}+a^{\dagger}e^{i\nu})^k,
\label{eq:rozwiniecie}
\end{equation}
in the rotating wave approximation and in an interaction picture the
Hamiltonian is approximated by the first five terms of the expansion
\begin{eqnarray}
H^{\mathrm{rwa}}_{\mathrm{int}} &=& \hbar \frac{\Omega}{2} \left\{
1-\frac{\eta^2}{2} + \frac{\eta^4}{8} + \left( -\eta^2 +
\frac{\eta^4}{2}\right)a^{\dagger}a \right. \nonumber\\ &+&
\left. \frac{\eta^4}{4}{a^{\dagger}}^2a^2 \right\}(\sigma^+ +
\sigma^-).
\label{rozwiniecie}
\end{eqnarray}
The terms in the first line in the above formula govern the free ion
evolution.  The last term in (\ref{rozwiniecie}) we associate with a
self-Kerr interaction Hamiltonian, up to the electronic state
operators.  Therefore, the unitary evolution operator resulting from
the nonlinear part of the Hamiltonian is given by
\begin{equation}
U(t) = e^{i \frac{\Omega \eta^4 t}{8} {a^{\dagger}}^2a^2 \, (\sigma^+
  + \sigma^-)}.
  \label{unitary-nonlinar}
\end{equation}
It allows for reading out the effective value of the evolution parameter
\begin{equation}
\tau_{\mathrm{eff}} = \frac{\Omega \eta^4 t}{4}.
\end{equation}
In the further discussion we neglect the electronic state evolution
since it may be made to take no part in the dynamics. To do this one
first needs to prepare the electronic state in an eigenstate of
$\sigma_x$ using a $\pi/2$ pulse tuned to the carrier transition, but
the pulse must not excite the vibrational degree of freedom in any
way. This can be done by making the pulse propagate orthogonal to the
vibrational axis.

Please note that by way of contrast to the method described in the
previous section this method is independent of the initial amplitude
$\alpha$.  The laser pulse duration required for the Kerr state
generation is the same for all values of the amplitude.  In other
words, the time required for generation e.g. $|\Psi(\alpha=2,
\tau)\rangle$ and $|\Psi(\alpha=5, \tau)\rangle$ is the same. However,
accessibility of the evolution parameter $\tau_{\mathrm{eff}}$ is
limited instead.  The greater Rabi frequency $\Omega$ and Lamb-Dicke
parameter value are the shorter pulse duration $t$ must be to obtain
$\tau_{\mathrm{eff}}$.  On the other hand, the ion has to remain
within the Lamb-Dicke regime in order to hold the expansion
(\ref{rozwiniecie}) true.

\begin{figure}[H]
\begin{center}
\scalebox{0.7}{\includegraphics{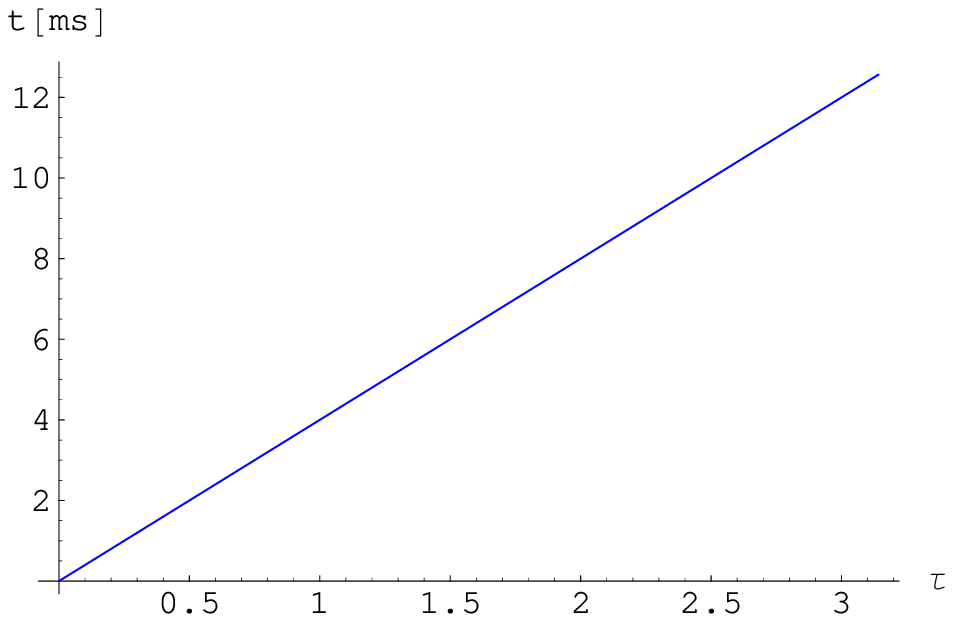}}
\scalebox{0.7}{\includegraphics{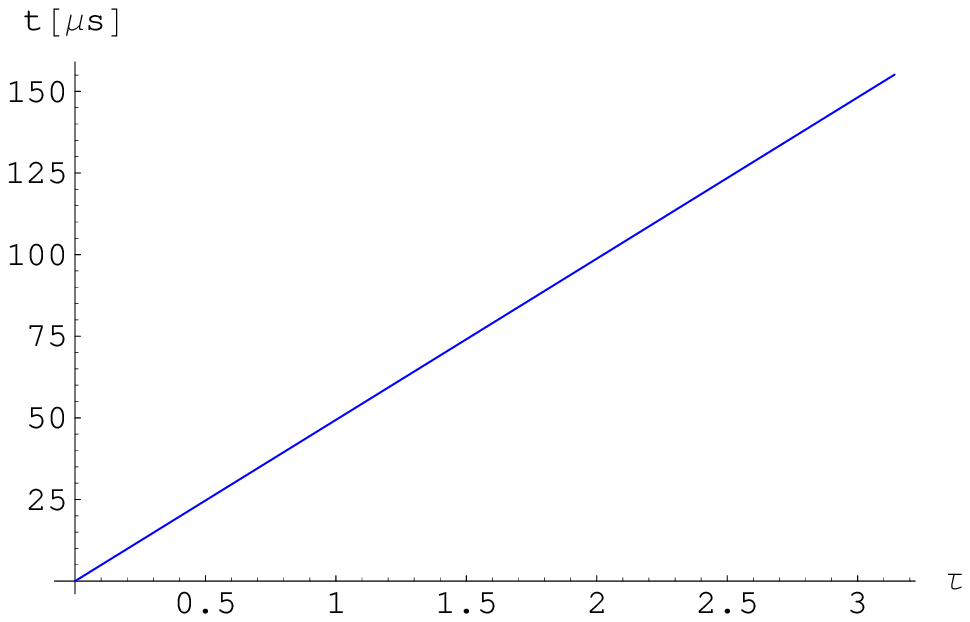}}\\
\scalebox{0.7}{\includegraphics{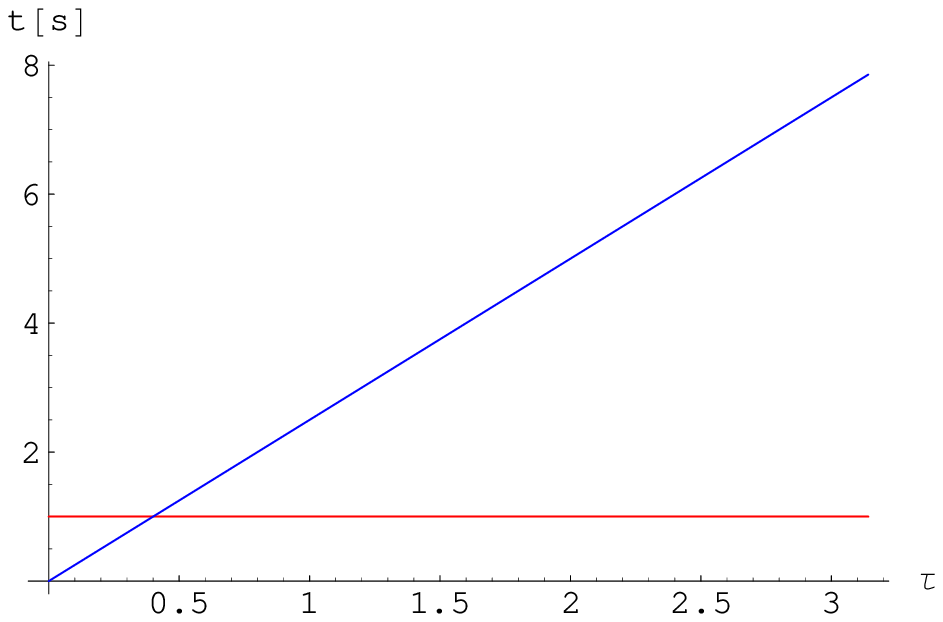}}
\scalebox{0.7}{\includegraphics{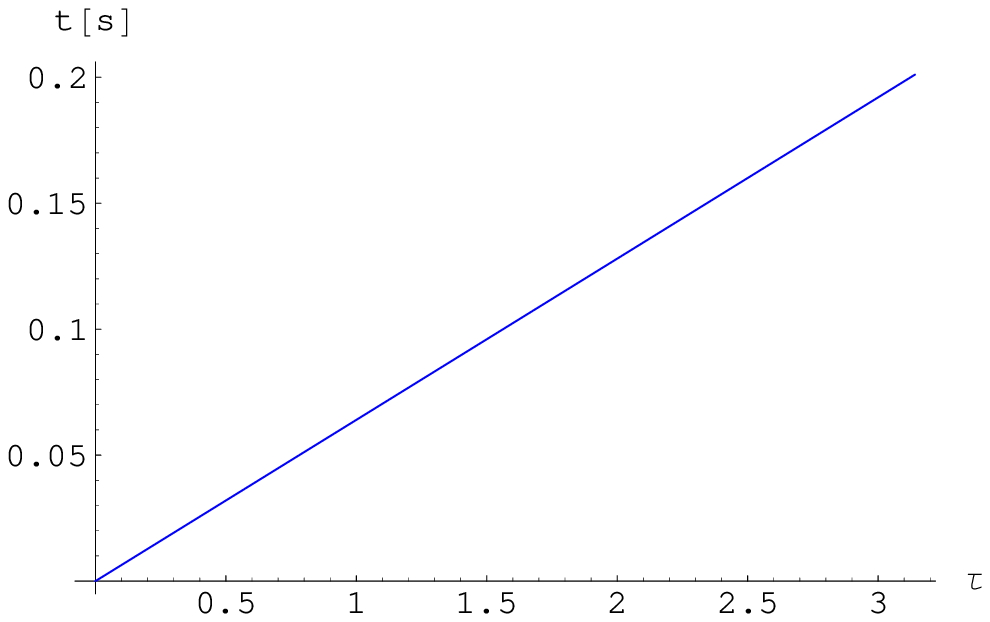}}
\end{center}
\caption{The time of pulse duration required for Kerr state
  $|\Psi(\alpha,\tau)\rangle$ generation for Lamb-Dicke parameter
  $\eta=0.1$ -- the top figure, $\eta=0.3$ -- the top middle figure,
  $\eta=0.02$ -- the bottom middle figure, $\eta=0.03$ -- the bottom
  figure.  The red horizontal line denotes time duration of
  $1\,\mathrm{s}$.  }
\label{Fig:one_pulse_duration}
\end{figure}

Figure \ref{Fig:one_pulse_duration} shows the time of the pulse
duration assuming its Rabi frequency $\Omega = 10 \, \mathrm{MHz}$ and
the Lamb-Dicke parameters $\eta=0.1$, $\eta=0.3$, $\eta=0.02$, and
$\eta=0.03$.  For the first two Lamb-Dicke parameters it is less than
$800 \, \mu\mathrm{s}$, which is within the coherence time of vibronic
ion state ($190\, \mathrm{ms}$), to obtain $\tau_{\mathrm{eff}}=\pi$
and therefore produce the cat state.  For the two remaining $\eta$
values the duration of the pulse is of order of tenth of second.

The duration of the pulse required for
$|\Psi(\alpha,\tau=0.04)\rangle$ generation is equal to
$t=0.16\,\mathrm{m}\mathrm{s}$ for $\eta=0.1$ and $t=1.98
\,\mu\mathrm{s}$ for $\eta=0.3$ ($t= 0.1\,\mathrm{s}$ for $\eta=0.02$
and $t=0.2\,\mu\mathrm{s}$ for $\eta=0.03$). For $\tau=0.04$ the first
negativities in the Wigner function become visible~\cite{PRA}.

This method allows for obtaining also the other coherent state
superpositions easily.  For example, for $\tau_{\mathrm{eff}} =
\frac{\pi}{3}$ one achieves superposition of six states and if
$\tau_{\mathrm{eff}} = \frac{\pi}{4}$ one achieves superposition of
four states (the compass state) \cite{OSID}.  For $\eta=0.3$ and
$|\Psi(\alpha,\tau=\pi/3)\rangle$ the pulse duration is equal to
$t=51.71\,\mu\mathrm{s}$, and for $|\Psi(\alpha,\tau=\pi/2)\rangle$ it
is equal to $t=77.57 \,\mu\mathrm{s}$.  These values are
experimentally feasible.

The one laser pulse method of $\chi^{(3)}$ non-Gaussian state
generation is limited due to introduced cut-off in the expansion
(\ref{eq:rozwiniecie}). It is valid only if we take the expansion for
a small parameter. To estimate the small parameter we approximate the
quantum operators by the classical amplitudes $a \simeq \alpha$
\begin{multline}
e^{i\eta(ae^{-i\nu}+a^{\dagger }e^{i\nu})} = \sum_{k=0}^{L}
\frac{(2i\eta|\alpha|)^k}{k!}\left(\frac{ae^{-i\nu}+a^{\dagger}e^{i\nu}}
     {2|\alpha|}\right)^k
\end{multline}
and read out the small parameter in the expansion to be $2\eta
\alpha$.  The above series is decreasing if $\eta \alpha <
\frac{1}{2}$.

This constraint is enough to ensure that the contribution from the
higher order terms is negligible.  For example, for $\alpha=5$ and
$\eta=0.1$, obeying $\alpha = \frac{1}{2\eta}$ ($\eta =0.09$ and
$\alpha < \frac{1}{2\eta}$) the coefficients
$\frac{(2\eta|\alpha|)^k}{k!}$ are equal to: $0.5$ ($0.405$) for
$k=2$, $0.041$ ($0.027$) for $k=4$, $0.0013$ ($0.0007$) for $k=6$. For
coefficients obeying $\alpha < \frac{1}{2\eta}$ the sixth order
coefficient is of two orders of magnitude smaller than the forth order
coefficient.

\section{Weak force detection}

We now show how the cat-state weak force detection protocol proposed
in Munro et al \cite{Munro2002} can be implemented in an ion trap
given the ability to engineer an approximate cat state as we have
described. It must be admitted that there is no compelling case to use
the vibrational motion of a trapped ion as a weak force
detector. However, the method we propose here would be a nice
demonstration of how non-Gaussian states can beat the standard quantum
limit for weak force detection.

Suppose one was able to prepare the vibrational degree of freedom in
the non-Gaussian state
\begin{equation}
|\psi_i\rangle=\frac{1}{\sqrt{2}}\left (e^{i\pi/4}|\alpha\rangle
+e^{-i\pi/4} |-\alpha\rangle\right )
\end{equation}
with the amplitude $\alpha$ real. When a weak force acts it can be
described by the action of the unitary displacement operator $D(i
\epsilon)=\exp(i\epsilon a^\dagger-i\epsilon a)$ acting on the initial
state $|\psi_i\rangle$ to give the output state $|\psi_o\rangle=D(i
\epsilon)|\psi_i\rangle$. Using the result that
\begin{equation}
D(i\epsilon)|\alpha\rangle=e^{i\mathrm{Im}(i
  \alpha\epsilon)}|\alpha+i\epsilon\rangle
\end{equation}
 we find that for $\epsilon \ll \alpha$, and $\alpha$ real that
 \begin{equation}
 |\psi_o\rangle=\cos(\pi/4+\alpha\epsilon)|+\rangle+i\sin(\pi/4+\alpha\epsilon)|-\rangle
\label{roatated}
 \end{equation}
 where the even and odd parity states are given by
 \begin{equation}
|\pm\rangle=\frac{1}{\sqrt{2}}\left (|\alpha\rangle\pm|-\alpha\rangle\right )
\end{equation}
In other words, the weak force is well approximated by a rotation in
the two dimensional parity subspace.  As in \cite{Munro2002} it then
follows that the minimum detectable force is then given by
\begin{equation}
\epsilon_{min}=\frac{1}{2\alpha}
\end{equation}
which beats the standard quantum limit by a factor of $(\alpha)^{-1}$.

Ion trap provide a simple way to reach this lower bound. Suppose that
after the weak force has acted, the total vibrational and electronic
state is given by Eq. (\ref{roatated}).  The first step is to make a
$\pi/2$ rotation of the electronic state, followed by the conditional
rotation
 \begin{equation}
 R=e^{-i\pi a^\dagger a\sigma_z}
 \end{equation}
 which can easily be done \cite{ion-parity}. Finally another $\pi/2$
 rotation gives the state
  \begin{equation}
 |\psi_o\rangle=\cos(\pi/4+\alpha\epsilon)|+\rangle|g\rangle+i\sin(\pi/4+\alpha\epsilon)|-\rangle|e\rangle.
 \end{equation}
The electronic state can now be readout, and the probability to find
the ion in, say, the excited state is
\begin{equation}
P_+=\frac{1}{2}\left (1-\sin(2\alpha\epsilon)\right ).
\end{equation}
Sampling this distribution gives the estimation for the force with a
minimum detectable force that is inversely proportional to
$\alpha$. For example, the minimum force required to shift the
interference distribution by one fringe is  $\epsilon_{min}=\pi/(4\alpha)$.

\section{Conclusions}

In the paper we have presented two experimentally feasible methods of
$\chi^{(3)}$ non-Gaussian states generation for light using a trapped
ion.  The first method is based on a well-known protocol (a series of
laser pulses) and allows for generation of an arbitrarily well
approximated non-Gaussian state.  The approximation is quantified by
three criteria based on Wigner function analysis.  This method is only
limited by technical parameters of the trap: the binding energy.

The second method enables an exact non-Gaussian state generation using
one laser pulse.  It is limited by the decoherence time of the
motional ion state.  However, adjusting the laser pulse and trap
parameters properly one is able to produce a cat state.

Based on the proposed protocols \cite{Zeng1994,Parkins1999,Pope2004}
we believe that non-Gaussian state transfer from ion motion to a light
beam is possible in a foreseeable future.  It will enable application
of a nontrivial quantum computing protocols based for example on
coherent states.  This is also a step towards exploring so far unknown
branch of quantum optics such as non-Gaussian states.

\section*{Acknowledgments}
M. S. acknowledges support by the Alexander von Humboldt foundation
and by MEN Grant No. N202 021 32/0700. M. S. also thanks I. Cirac for
the discussion.

\end{document}